\documentclass[10pt,conference]{IEEEtran}
\usepackage[pdftex]{graphicx}
\usepackage{listings}
\usepackage{multirow}
\begin{document}

\title{Efficient and reliable network tomography in heterogeneous networks using BitTorrent broadcasts and clustering algorithms}


\author{
    \IEEEauthorblockN{Kiril Dichev\IEEEauthorrefmark{1}, Fergal Reid\IEEEauthorrefmark{1}\IEEEauthorrefmark{2}, Alexey Lastovetsky\IEEEauthorrefmark{1}}
    \IEEEauthorblockA{\IEEEauthorrefmark{1}School of Computer Science and Informatics
\IEEEauthorrefmark{2}Clique Research Cluster\\
University College Dublin, Belfield, Dublin 4,\\Ireland
    \\\\ Kiril.Dichev@ucdconnect.ie\\
	Fergal.Reid@gmail.com \\
	Alexey.Lastovetsky@ucd.ie}
}


\maketitle

\begin{abstract}
In the area of network performance and discovery, network tomography focuses on reconstructing network properties using only end-to-end measurements at the application layer. One challenging problem in network tomography is reconstructing available bandwidth along all links during multiple source/multiple destination transmissions. The traditional measurement procedures used for bandwidth tomography are extremely time consuming. We propose a novel solution to this problem. Our method counts the fragments exchanged during a BitTorrent broadcast. While this measurement has a high level of randomness, it can be obtained very efficiently, and aggregated into a reliable metric. This data is then analyzed with state-of-the-art algorithms, which reliably reconstruct logical clusters of nodes inter-connected by high bandwidth, as well as bottlenecks between these logical clusters. Our experiments demonstrate that the proposed two-phase approach efficiently solves the presented problem for a number of settings on a complex grid infrastructure.
\end{abstract}
\begin{IEEEkeywords}
Network tomography, BitTorrent, clustering, bandwidth, bottleneck link
\end{IEEEkeywords}

\section{Introduction and Related Work}
The properties of the underlying network play a central role in the performance of all distributed and parallel applications which rely on communication. When these properties are taken into account, communication can be optimized. In the Message Passing Library (MPI), every collective operation can profit through topology awareness, particularly in heterogeneous networks. A large body of research has been done in this direction for various protocols and networks, including but not limited to \cite{Gabriel98,Kielmann2000, Karonis2002-MPICH-G2,Coti2009, Subramoni2011} .
Existing work performs topology-aware collective operations using knowledge of a pre-defined partition clustering of the network, and finds that these topology-aware collectives substantially outperform topology-agnostic methods.
In this work, our goal is to provide a versatile automated method to deduce a partitioning of the network into logical bandwidth clusters. 
We aim to provide a method that is efficient in its network measurement, and good at finding network bottlenecks under conditions of high load.
This method would allow easy topology aware communication on large highly utilized heterogeneous networks, which are becoming an increasingly important domain for distributed computation.

There are two main ways of incorporating knowledge about the heterogeneity of the network into communication algorithms - by providing such knowledge manually or automatically. Past MPI implementations for Grid infrastructures and other wide-area networks \cite{Gabriel98,Kielmann2000, Karonis2002-MPICH-G2} have used each of these ways of incorporating such knowledge. Fully automatic approaches can be subdivided into `intra-node' and `inter-node' approaches. For inter-node approaches, network discovery typically involves some form of communication model. There are many examples of this approach in high-performance computing and early work includes \cite{Bhat1999,Hatta2000}. For intra-node automatic approaches, which are gaining popularity today due to many-core nodes, recent work includes \cite{Graham2008}. 

\begin{centering}
\begin{figure}
\includegraphics[width=3.5in]{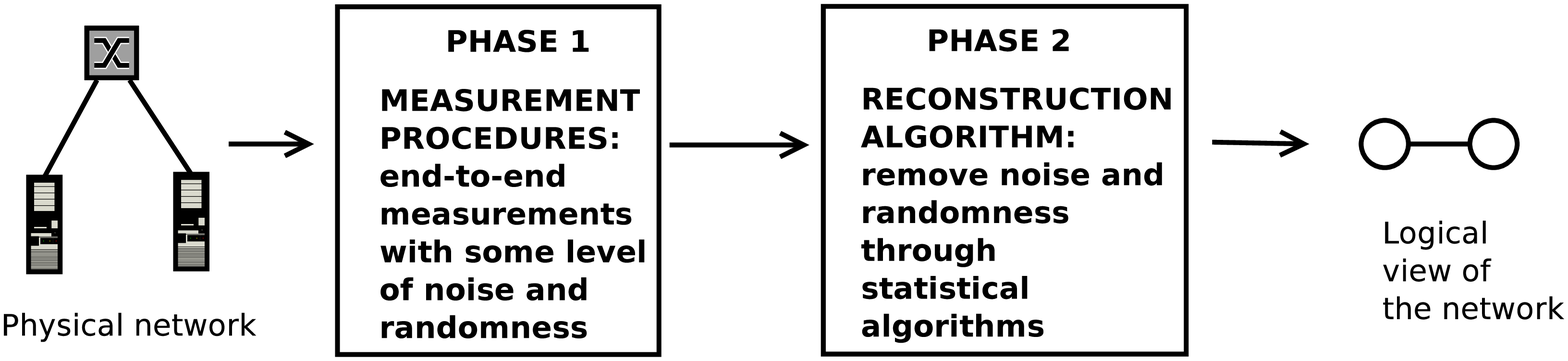}
\caption{Overview of network tomography.}
\label{fig:tomography}
\end{figure}
\end{centering}

However, the existing automated approaches in high-performance computing do not generally capture all relevant network properties. 
Examples of properties that existing approaches are poor at capturing include bottleneck links; as while often not visible in isolated point-to-point communication, these bottlenecks appear under conditions of particularly intense collective communication. 
We look for possible solutions to these peak-bandwidth measurement issues in distributed computing. 
The challenges of network discovery in this area gave rise to an interesting sub-field of research in the late 90s, called `network tomography'.
Castro \cite{Castro2004} provides a detailed overview. 
The general goal of network tomography is to reconstruct the logical topology of the network in two phases, as depicted in Fig. \ref{fig:tomography}. 
The first phase involves only end-to-end measurements of the network. 
Based on how these measurements are performed - whether actively or passively - we can talk of passive network tomography (e.g. \cite{Padmanabhan2002}) or active network tomography (e.g. \cite{Lawrence2007}). 
After measurement data is collected, the second phase of the process always involves the use of statistical methods to reconstruct the logical view of the network.  
While many metrics can be used, a number of metrics are particularly relevant from the user perspective, including loss rate, delay and bandwidth. 
Network tomography can be used both in wide-area networks and in networks consisting of clusters of clusters. 
It is mostly used in networks with some level of heterogeneity and/or hierarchy. 

The main contribution of this work is in the area of active network tomography with a bandwidth-related metric. 
Our focus is ``multiple source multiple destination'' communication.
An example of such communication would be heavy bulk transfer between all network peers (e.g. in all-to-all communication). 
The closest related work to ours are \cite{Bobelin2008} and \cite{Legrand2002}. 
Both works reconstruct the logical topology of the network using bandwidth as a metric. 
While \cite{Legrand2002} infers a qualitative view of the network, \cite{Bobelin2008} infers a more quantitative view, including labeling of actual achievable bandwidth. 

\begin{centering}
\begin{figure}
\includegraphics[width=3.5in]{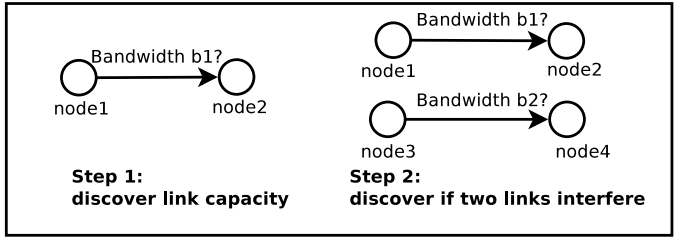}
\caption{Traditional measurement of bandwidth in tomography.}
\label{fig:trad-metric}
\end{figure}
\end{centering}

The most time-consuming phase in existing approaches to active network tomography on bandwidth is the measurement phase. 
The measurement procedures used to find the available bandwidth and/or bottlenecks between communicating peers are generally similar across all tomography methods. 
We show the two essential steps that are generally involved in Fig. \ref{fig:trad-metric}. 

\begin{centering}
\begin{figure}
\includegraphics[width=3.5in]{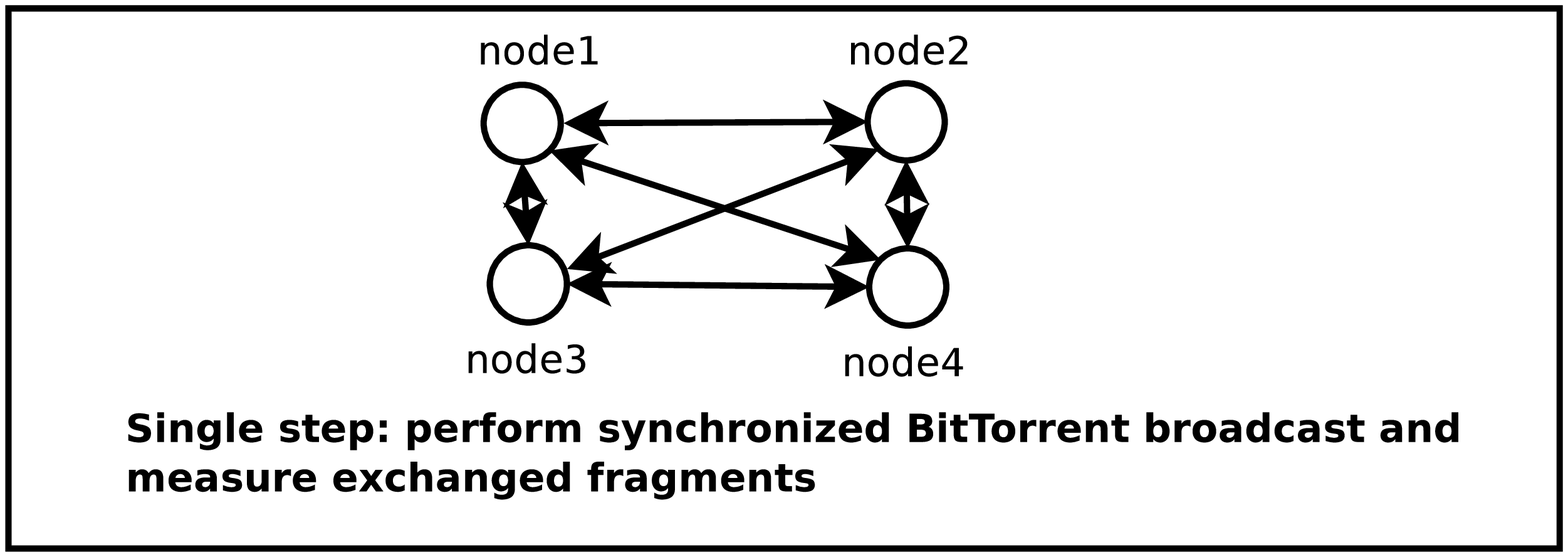}
\caption{Alternative bandwidth-related measurement as proposed in this work.}
\label{fig:bt-metric}
\end{figure}
\end{centering}

Conceptually, in the first step, an intense communication is established between a pair of nodes until the link capacity between them is reached.
Then, a new pair of intensely communicating nodes is introduced, and the bandwidth of the first link is reexamined. 
If no change in bandwidth is observed, then the links are probably independent - more pairs communicating in parallel could unveil a bottleneck at a later point. 
But if the bandwidth of the node pairs under examination decreases, then it is clear that they share the same physical (and logical) link. 
Following this procedure, experiments are performed until the entire network is reconstructed. 
While intuitive, this approach is very expensive. 
The measurement procedures have polynomial complexity, even after some optimizations using heuristics or parallelism. 
Indeed, this complexity makes it infeasible to perform bandwidth-related network tomography on large-scale computer networks. Both  \cite{Bobelin2008} and \cite{Legrand2002} concede this issue and resort to running simulations using SimGrid \cite{simgrid}. \cite{Legrand2002} only attempts real experiments on a small scale. Even simplified measurement procedures require approximately one hour to run with only 20 nodes.

In our work, we introduce a novel approach to multiple source / multiple destination network tomography, which differs from traditional approaches described in Fig. \ref{fig:tomography} in both the measurement procedure and the analysis method as follows:

\begin{itemize}
\item Phase 1:
We use the BitTorrent \cite{Cohen2003} protocol as the tool for our measurements. The measurement metric is the number of exchanged fragments between peers during a synchronized BitTorrent broadcast (see Fig. \ref{fig:bt-metric}).  
\item Phase 2: 
In the second phase, we use a popular implementation of a `modularity' based network clustering algorithm \cite{blondel2008fast} \cite{newman2004finding} to perform analysis of the measured data. 
We describe this process, and choice of clustering algorithm, as well as its operation, in detail in Section \ref{ModularityBasedClustering}.
\end{itemize}

We are not aware of previous work in this area using the type of measurement procedure we employ, nor are we aware of any coupling of this approach with a clustering algorithm in the manner which we present. 

The intuition behind the metric is that when using a number of parallel connections, more data will be naturally transferred through the links with higher bandwidth. 
This idea has been demonstrated even before the advent of protocols like BitTorrent. 
For example, \cite{Rodriguez2002} demonstrated that a client using a number of parallel TCP connections to servers with different upload rates will download a file with a rate approaching the fastest upload rate. 
Indeed, the BitTorrent protocol uses a number of parallel connections to exploit this network feature. 
The proposed measurement departs from existing approaches in the field not only in the metric, but also in its efficiency - a single synchronized BitTorrent broadcast can often capture the behavior of a very large number of links. 
Indeed, every broadcast of a large message to many peers has a total communication time linear in the message size. 
The number of peers does not reduce the download rate due to the pipelining and scalability of the BitTorrent algorithm. 
For detailed performance analysis of BitTorrent see for example \cite{Xia2010,Izal2004}.

Surprisingly, we find only one other work \cite{Liu2006} which investigates BitTorrent tomography. However, this work states that BitTorrent traffic is immeasurable on a large scale. The authors do not list any technical issues with this approach, but argue that it is unlikely for an instrumented BitTorrent client to be used by a large user community. They replace BitTorrent profiling with an alternative algorithm, which they use in conjunction with a simulation program.
There is also work which proceeds in the opposite direction - optimizing the performance of the BitTorrent protocol, given knowledge of the network topology (e.g. \cite{Ren2010}).

In order to validate our tomography method we perform a set of experiments on real networks using a grid infrastructure.
Although we focus on bulk transfers of large data, our measurement approach is efficient, because it does not perform exhaustive measurements. 
The measurement procedure, which consists of several iterations of BitTorrent broadcasts, captures the flow of large data volumes across the entire network in each iteration. 
With just a few iterations, we find that sufficient information can be collected to infer important network properties.
This information can be used in many data-intensive communication operations. 
For example, if we want to efficiently schedule an all-to-all operation, we do not need to label the achievable bandwidth along all fast and slow links in the network.
Instead, the only requirement is a logical clustering of nodes according to their bandwidth. Thus, if there is a bottleneck link between nodes, a correct clustering algorithm should place them in different logical clusters.

We find that our method efficiently and reliably clusters nodes with regard to their bandwidth when all nodes are involved in collective communication. 
We use the algorithm successfully both for experiments separating compute clusters within a site, and between sites, on the Grid'5000 infrastructure.

The paper is structured as follows: Section \ref{metric-section} presents our metric and measurement procedures. Section \ref{ClusteringMethod} presents the clustering algorithm we use. In Section \ref{ExperimentalResults} we present results from a range of experiments. We conclude the paper in Section \ref{Conclusion}.

\section{BitTorrent Broadcasts and ``Received Fragments Per Peer'' as a Metric}
\label{metric-section}
\subsection{Definition and Use}
First we define the metric formally. We observe the communication network as a directed graph G=(V,E). Throughout this paper, we refer to a BitTorrent broadcast as a \textit{fully synchronized instrumented execution of BitTorrent clients until all clients have downloaded a file}. A file of size M is distributed as $\frac{M}{16KB}$ fragments of 16KB to all nodes v $\in$ V using the BitTorrent protocol. If $v_1 \rightarrow_i v_2$ denotes the number of fragments sent directly from $v_1$ to $v_2$ within broadcast operation i, then we define the metric w per edge e for one run as 
\begin{equation}
w(e) = v_1 \rightarrow_1 v_2 + v_2 \rightarrow_1 v_1
\end{equation}
with $e=(v_1,v_2)$.
Since performing more iterations significantly increases the accuracy of the metric, for n iterations we simply state

\begin{equation}
w(e) = \frac{\sum_{i=1}^n{v_1 \rightarrow_i v_2 + v_2 \rightarrow_i v_1}}{n}
\end{equation}
with $e=(v_1,v_2)$.

In this work, the size of a file used in a single BitTorrent broadcast is chosen to be 239 MB. This choice is completely arbitrary and is driven by practical observations that a single broadcast then takes around 20 seconds for different numbers of nodes (see Section \ref{efficiency-of-the-metric} for details). We find this to be a reasonable amount of time for a single broadcast, that often provides good bandwidth information for many of the available links. Profiling the precise number of fragments exchanged gives the following information:  in each BitTorrent broadcast, exactly 15259 fragments of 16384 bytes each are received by all participating nodes, following a dynamic pattern each time.

We have instrumented the original Python version of the BitTorrent client written by Bram Cohen and available in most Linux distributions. We introduce efficient profiling of the arriving data as follows: At the reception of each data fragment, a counter is incremented associated with the sending peer using a hash table of counters. At the end of a run, all peers have a record of the source peers and the number of fragments they received from each peer.

\begin{centering}
\begin{figure}
\includegraphics[width=3.5in]{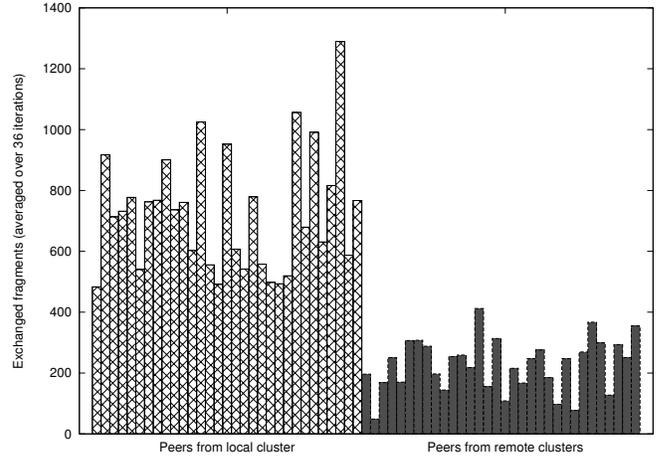}
\caption{Measured metric values for all edges to a randomly fixed node (36 iterations). On the left are edges to local cluster nodes, on the right are edges to remote nodes.}
\label{fig:sample-histogram}
\end{figure}
\end{centering}

As an example, in a broadcast operation involving 64 nodes on one site, we display measurements for a randomly chosen node in Fig. \ref{fig:sample-histogram}. The bars represent the metric as defined above for 36 iterations for all edges which include the fixed node. Since the results involve many iterations, the chosen node exchanges fragments with all 63 peers. For clarity, we have grouped on the left side the metric values for the 31 peers in the local cluster, and grouped the values for the 32 remote nodes on the right side. 

We will discuss the main characteristics of this measurement, and explain how it differs to classic bandwidth measurements in the following sections.

\subsection{Efficiency of the metric}
\label{efficiency-of-the-metric}

The main strength of our method is that it takes only a single broadcast of a large message per run to collect data on a large subset of all possible peer-to-peer connections. In our setup, the observed complexity of BitTorrent broadcasts is $O(M)$ -- linear in the message size M. We verified experimentally that as we alter the number of nodes, the BitTorrent broadcast requires nearly constant time. According to practices from high-performance computing, our reference time for the completion of a BitTorrent broadcast is the maximum download completion time of all the BitTorrent clients, which we start synchronously. 
For 32, 64 and 128 nodes, the broadcast of the 239 MB large message takes about 20 seconds on the Grid'5000 infrastructure, even when the nodes are spread across 4 sites. %
Related work \cite{Izal2004} also suggests that a high download rate can be sustained for very large peer numbers; the number of participating peers in such experiments typically does not alter the estimated time of $O(M)$ for all peers to download the file. Other work \cite{Zissimos_gridtorrent:optimizing} also demonstrates that the BitTorrent protocol is competitive with client/server architectures in its peak download rate.

We now give a short overview of the complexity of the network tomography algorithms we previously mentioned. 
Each step in these algorithms -- shown in Fig. \ref{fig:trad-metric} -- is very time consuming. 
First, every link has to be saturated until the maximum bandwidth on that link is reached. 
This is a costly operation which incurs heavy network overhead. 
The second challenge consists of probing the link bandwidth in parallel for multiple links.
This process is repeated until all nodes have been sufficiently tested. For example, \cite{Bobelin2008} performs such tests only with at most triplets of nodes. It is stated that triplets are sufficient as long as the single-link experiments can reach the maximum capacity. Even with this assumption, all possible triplets need to be tested in the worst case. This step is performed since it is assumed that there is no a priori knowledge of the topology of the network. The observed complexity of the algorithm in this case is $O(N^3)$, in $N$ the number of nodes.

The algorithm proposed by \cite{Legrand2002}, on the other hand, tests pairs incrementally, fully in parallel and without limiting the maximum number of tested links at a time.
In specific cases, where no interference of links is observed, the complexity is estimated at $O(N^2)$. 
The only empirical experiments performed are for networks of 20 nodes, and these take about one hour to complete.

\subsection{Level of randomness with single runs using the metric}
If we examine the volume of exchanged data shown in Fig. \ref{fig:sample-histogram} over a number of iterations, we notice that a total of 22533 fragments are exchanged with local cluster nodes, and 6337 fragments are exchanged with remote nodes. 
This is a clear indication that with BitTorrent broadcasts, data flows with a preference for high bandwidth links. 
Furthermore, we observe this phenomenon quite reliably in our experimental data.

\begin{figure}
\includegraphics[width=3.5in]{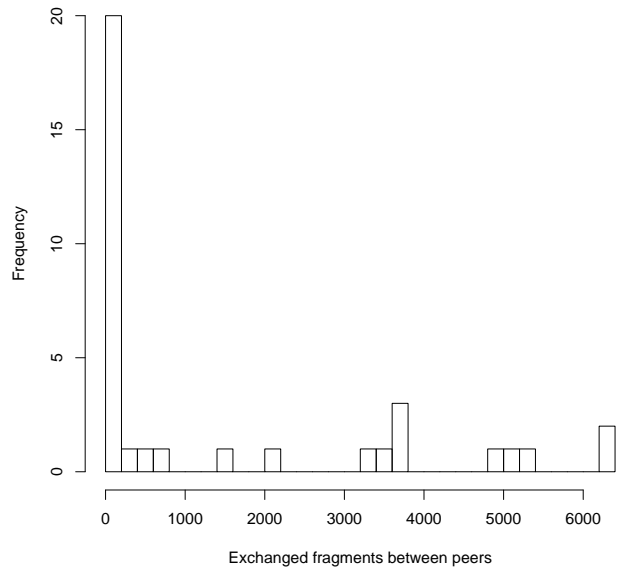}
\caption{Distribution of measurement of edge bandwidth w(e) for a fixed edge e for 36 independent iterations.}
\label{fig:histogram}
\end{figure}

We have previously defined a single run of our metric as transmitting a single file which takes approximately 20 seconds.
As the operation of the BitTorrent protocol is stochastic, and the data transferred across each link varies from run to run, it is important to attempt to characterize the accuracy of the metric we have defined, when a single run is performed. 
Thus, we now observe how the metric fluctuates using one Grid'5000 site (Bordeaux). 
We focus on an edge between 2 nodes randomly chosen from within a cluster. 
Each run measures the metric $w(e)$ independently (no aggregation is used). 
Fig. \ref{fig:histogram} shows the distribution of $w(e)$ along the fixed edge, over 36 runs. 
In 23 of the 36 runs, the two peers do not exchange any data with each other. 
In the remainder of the runs, the exchanged data varies between 3 and 6304 fragments. 
This distribution shows that the variance is very high. 
For comparison, when running the well known NetPIPE tool \cite{Snell1996} to establish the maximum achievable bandwidth along the link between two peer nodes on the same compute cluster used above, the variance is very low and the distribution is dense around 890 Mbps.

Fig. \ref{fig:histogram} suggests that while inexpensive to compute, the metric is very variable for single runs. 
With this level of measurement noise and randomness, a good analysis technique will be needed to extract meaningful data from these measurements.%
Yet one important consideration is that our analysis method does not consider each link's bandwidth in isolation, as in previous approaches, and this eases to some extent the requirements for the measurement step.

Before discussing how we will aggregate and analyze data from this metric to form a reliable view of the network, we will briefly discuss the BitTorrent properties which are responsible for the variance and high degree of randomness between single runs of the metric:
\begin{itemize}
\item Initially, BitTorrent clients \emph{randomly} choose their initial peers (adjustments in the peer selection are part of the protocol for longer runs).
\item BitTorrent internally limits the number of parallel uploads to 4, and this indirectly limits the number of parallel downloads.
\item Another protocol feature is that the number of total peers is limited to 35. This means that for larger numbers of nodes and a single broadcast, measurements using this protocol will not provide a complete graph - only a subset of possible connections will be measured. One solution to this problem is to aggregate the measurements over a number of BitTorrent broadcasts, as we shall see.
\item Using a BitTorrent broadcast operation means that nodes which are better connected to the `root node' are more likely to receive more fragments from the root. This is simply due to the asymmetric way data flows in a broadcast operation as compared to, for example, an all-to-all transmission. However, in our experiments this was never an issue during the reconstruction and analysis of our networks. If this affects results in some cases, a simple solution is using different root nodes over a number of runs.
\end{itemize}

These are characteristics of the protocol, which, while important for transmission efficiency and reliability, increase the variance of our measurements, and could make network reconstruction hard.
However, it will become evident in the following sections that despite the high degree of inherent randomness, this metric can easily be made reliable through simple iteration, especially when the measurements are analyzed using a clustering algorithm which operates on the observed network as a whole.

\subsection{Improving the accuracy through iteration of BitTorrent broadcasts}
While a single broadcast measurement has a high level of noise and randomness, aggregating data over a number of iterations resolves these issues. A positive property of the used approach is that each new iteration potentially improves the accuracy of the metric on a global scale -- i.e. for all edges. 
The previous approaches presented can not address this, and are restricted to local experiments on a small subset of nodes. 
In order to quantify the number of iterations needed to improve our accuracy, the key questions are:
\begin{itemize}
\item How close is the single run data to an ``ideal'' representation of the peer-to-peer bandwidth when performing bulk data transfers?
\item How fast does the aggregated data over a number of runs converge to the ``ideal'' representation?
\end{itemize}

We address these questions in an end-to-end manner, by quantitatively evaluating the performance of the entire system which uses these measurements. 
We make experimental observations as follows: after obtaining measurements at each run, we use a clustering metric -- presented in detail in Section \ref{comparing-network-clusterings} -- to assess the quality of measurements against a ground truth. %
Our experimental work (Section \ref{ExperimentalResults}) and in particular Figure \ref{fig:NMIresults} shows our data aggregation quickly converges to the ideal.
These questions could also be addressed by an analytical approach, but there are a number of challenges with this: the implementation of BitTorrent is not trivial to analyze, and porting it to a simulation environment is a complex engineering challenge. 
\section{Clustering method}
\label{ClusteringMethod}

The previous section has shown that the used measurement approach can be very efficient, but that there is a significant level of randomness in the data gathering process. 
This would seem to pose one significant challenge for a clustering algorithm.

A second challenge is that we want to provide as little a priori information to our tomography method as possible, in order to increase the range of possible application scenarios.
Specifically, we do not want to specify the number of logical clusters into which to partition the observed network.
We want to be able to deploy this method in an automated fashion, in real world application domains where very little information on the underlying topology is available. 

\subsection{Modularity based clustering}
\label{ModularityBasedClustering}
We apply a technique from modern network analysis to the problem of identifying logical clusters in the network.
We use the modularity function of Newman and Girvan \cite{newman2004finding} to identify sets of nodes which are more densely interconnected than the general level of interconnection in the network.

The modularity method is defined by the following objective \cite{newman2004finding}:
\begin{equation}
Q = \sum _{i}\left( e_{ii}-a_{i}^{2}\right) = Tr(e) -\left\| e^{2}\right\|
\end{equation}

which compares, for a given clustering, the proportion of network edges that are intra-cluster $e_{ii}$, for each cluster $i$, against the proportion that would be intra-cluster in a randomized model of the same network.
As described by Newman and Girvan: ``This quantity measures the fraction of the edges in the network that connect vertices of the same type (i.e., within community edges) minus the expected value of the same quantity in a network with the same community divisions but random connections between the vertices.''

We use a weighted version of this same objective, which will have a high value for clusters of nodes that have a high internal weight.
This objective has been applied in a wide range of domains, including finding communities of users in social networks, finding highly connected communication groups in telecoms networks, and many other related application problems.
As our objective is to find a partition of the network, into dense non-overlapping clusters, and in particular as we do not wish to specify beforehand the number of logical clusters to find, this objective function is appropriate.
In addition, our empirical results show it is effective at recovering the ground truth clusters as part of our tomography approach, as we discuss in Section \ref{ExperimentalResults}.
 
\subsection{Fast Louvain method}
Many different algorithms have been developed to optimize the modularity objective function. These algorithms improve on the original methods provided and are designed to work in practical settings and on large scale networks.
One of the most successful and widely used methods is that of Blondel et al. \cite{blondel2008fast}, known as the Louvain method.
This algorithm was originally developed and applied to large mobile telecommunications networks, in order to uncover clusters of frequently communicating users, and social communities; the authors found that they could uncover many levels of hierarchical organizational structure within the communications network.

While no meaningful close form complexity of this heuristic implementation is currently available, its fast runtime in practice, and ability to scale to large datasets, such as telecoms networks with millions of nodes, make this modularity optimization algorithm suitable for our purposes.

\subsection{Cluster visualization}
\label{cluster-visualisation}

As is shown visually in Figures \ref{neato:b}, \ref{neato:bt}, \ref{neato:gt}, \ref{neato:bgt} and \ref{neato:bgtl}, the application of an energy minimizing spring layout on the networks we have constructed often produces logical clusters that correspond to the logical clusters in the underlying computer network.
The nodes in these diagrams are represented as different shapes depending on the ground truth cluster to which they belong.
The exact details of how the ground truth is produced from the physical network topology will be discussed in Section \ref{ExperimentalResults}.
We layout these networks using a force directed implementation of the Kamada Kawai algorithm \cite{kamada1989algorithm} in the `Graphviz' software package \cite{ellson2004graphviz}, making the length of edges between nodes inversely proportional to the edge weight.
While we use all the measured edges in our layout algorithms, for clarity of presentation in these diagrams we only render the edges which are in the top 50\% of network edges by weight.
It can clearly be seen that the ground truth clusters provided match with the visually identifiable clusters of nodes formed by the force directed layout.
The fact that force directed layout works visually well on this network representation hints that a graph clustering method will be a suitable choice for a clustering which is predictive of the ground truth.

The work of Noack \cite{noack2009modularity} has shown an equivalence between modularity based network partitioning approaches, and the clusters formed by particular types of force directed layout.
The types of force directed layout discussed does not include the Kamada Kawai algorithm we use; still this does provide an indication that algorithmic clustering approaches will be successful on this problem.

\subsection{Other features}
The fast modularity maximization algorithm of \cite{blondel2008fast} produces a dendrogram of hierarchical clusters.
We do not use this dendrogram in this work; instead, we take the cut of the dendrogram at the point that yields the highest modularity value of the resulting partitions.
This results in only a single level of partitioning.
This is suitable for our purposes, as the ground truths that we use are partitions of the network; while the underlying networks themselves may be hierarchical, the ground truths are non-hierarchical in nature.
However, there is potential for extending this approach in future, and revealing structure of a more hierarchical nature.

Good et al. \cite{good2010performance} performed analysis of the modularity objective function, in a variety of practical contexts, and concluded that the optimization surface is often bumpy, and often lacks a clear global maximum in empirical settings; however, we find that this widely used community finding algorithm produces results that work well in this particular application domain.
Further, we find that repeated iterations of the optimization algorithm find results that are consistent with those presented in this paper; on the experimental networks we have examined, the algorithm seems to consistently converge to results that are in high agreement with our ground truth.

As an aside, we also attempted to perform these experiments, with another modern clustering algorithm, Infomap \cite{rosvall2008maps}, which is based on compressing random walks through the network, and finds communities which correspond to the areas of a network that a random walk would get `stuck' in. However, we find that this method does not perform as well as modularity based clustering for this particular problem.

\subsection{Comparing network clusterings}
\label{comparing-network-clusterings}
While the visualizations intuitively show a relationship between the ground truth partitions, and the clustering of the measured network, in order to quantitatively evaluate the success of our method, and in order to potentially evaluate it on networks too large to visualize, a numerical measure of cluster accuracy is necessary.

Many various methods for comparing set assignment exist.
In the domain of network community finding, a frequently used measure of comparison between a ground truth clustering, and an algorithmically provided clustering, is the informational theoretic measure of the Normalized Mutual Information between the two.
For convenience, and to enable the future extension of our work to situations where the ground truth overlaps, we use the overlapping NMI method of \cite{lancichinetti2009detecting}. 
This method is capable of calculating the NMI between a set of communities which overlap, as well as a set of network partitions.
This widely used method enables us to compare our clustering against the ground truth.
It ranges from 0 to 1, where 1 denotes perfect agreement of the found clustering with the ground truth.
We note that there are several improvements on this NMI method; we have also investigated the results of some of these, and observed consistent results; as such we report scores only for the popular NMI method of \cite{lancichinetti2009detecting}.

\section{Experimental results}
\label{ExperimentalResults}
\subsection{Introduction to experimental setup}
\label{ExperimentalResultsIntro}
In network tomography, an algorithm performs well if the reconstruction of the network is correct with regard to the dynamic bandwidth properties of the network. 
The purpose of the tomography method is to uncover these properties.
In practice, the relationship between these dynamic properties, and the physical structure of the network topology, is often complex.
However, in order to evaluate our method, we use the physical structure of the network topology, including information about how network hardware connects compute clusters within physical sites, and information on the speed of the inter-site links, to form a ground truth dataset.

\begin{figure}
\includegraphics[width=3.5in]{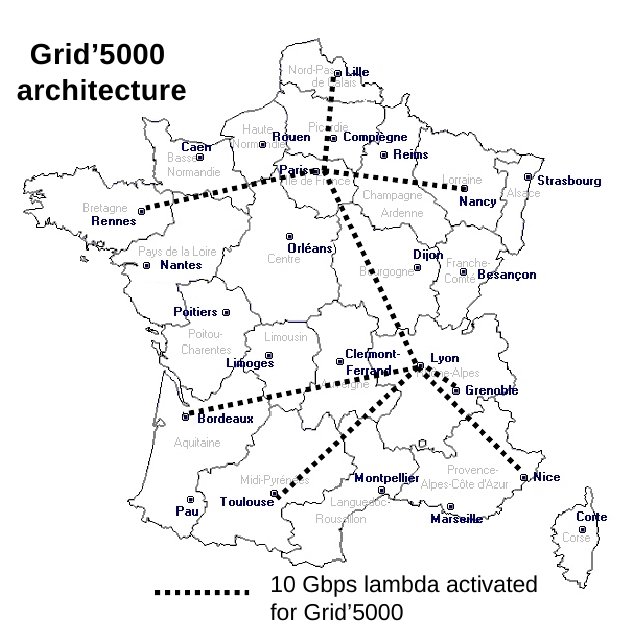}
\caption{The Renater infrastructure as presented in \cite{site:grid5000-networking}}
\label{fig:renater}
\end{figure}
We perform our experiments on the Grid'5000 infrastructure in France. Nine sites are interconnected using the Renater network (Fig. \ref{fig:renater}) providing high bandwidth optical fiber. Within each site, there are differing technologies, hierarchies and clusters. For this work, only the Ethernet network within sites as well as the Renater network between sites is used.

The a priori knowledge of the network, which is independent of the network tomography algorithm, is very important in this work. This knowledge provides our ground truth which we use to evaluate the found clustering. We have ground truth information about multiple aspects of the system:
\begin{itemize}
\item The communication between sites has similar properties - it uses the Renater infrastructure. While it provides very high inter-site bandwidth, it is reasonable to assume it will not outperform local Ethernet communication. Experiments using NetPIPE confirm this assumption - for example, the maximum bandwidth achieved between nodes on Bordeaux and Toulouse is around 787 Mbps - compared to 890 Mbps achieved within Ethernet clusters.

\item Within a Grid'5000 site, intra-site communication is complex. 
Physical hardware information is typically provided by online documentation available at \cite{site:grid5000-networking}. 
However, transient network anomalies can arise when observing the network behavior (e.g. bandwidth bottlenecks, availability of multiple Ethernet interfaces, hardware changes), and so the â'authoritative ground truth clustering is generally best provided by the site administrator.

\end{itemize}

When we use a setup spanning multiple sites, we assume the clustering should subdivide the network into separate logical clusters, each cluster corresponding to a single site. If we evaluate our method on a single site -- which we do for the Bordeaux network -- we generate our ground truth using the available information about the structure of the physical topology in that site. We discuss these specifics in each of our experiments in turn.

\begin{figure}
\includegraphics[width=3.5in]{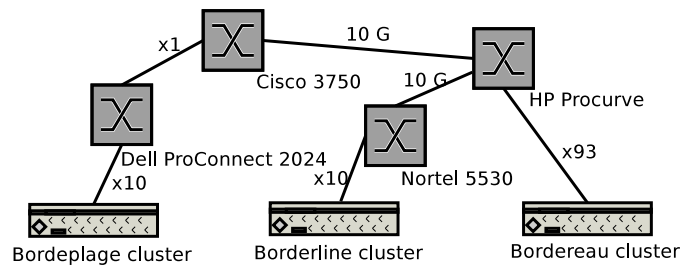}
\caption{Ethernet network at the 3 used clusters in the single site Bordeaux}
\label{fig:bordeaux-topology}
\end{figure}
\begin{figure}
\includegraphics[width=3.50in]{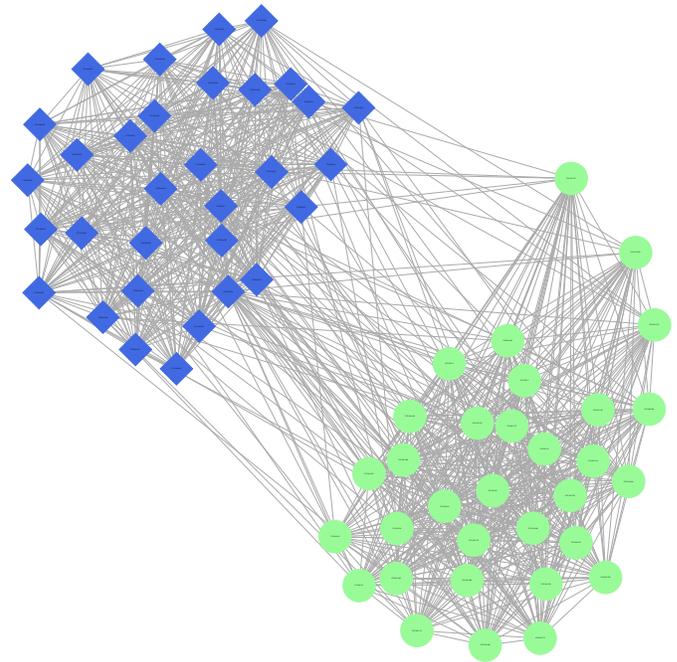}
\caption{Applying Kamada-Kawai layout (using the Graphviz' `Neato' tool) to dataset `B', the Bordeaux site. The configuration used has 64 nodes, divided between 3 physical compute clusters. These 3 physical compute clusters give rise to only 2 logical network clusters, as there is a fast link between the `Bordereau' and `Borderline' physical clusters. The shape and color of each node rendered reflects the labelling of the ground truth cluster it is in.  We render only the edges in the top half of all edges, by weight. While the graph is too dense to visually make out any structure due to edge weight, it is clear that the layout algorithm is clustering nodes according to their ground truth. This provides grounds for expecting a graph clustering algorithm to find these clusters.}
\label{neato:b}
\end{figure}

\subsection{One site experiments}
In Fig. \ref{fig:bordeaux-topology}, we display a partial view of the network topology in Bordeaux, excluding the connections to the external network and the Myrinet / Infiniband network. One important realization is that even when provided with an explicit diagram of the network, it still is not obvious where the bottlenecks and the strong links are in terms of achievable bandwidth. The site administrator clarified that the significant bottleneck is the link between the Dell and Cisco switches, which only provides a single 1 Gigabit Ethernet connection. Note that the link is only a bottleneck for the ``multiple source multiple destination'' scenario we are addressing in this work.

\subsubsection{2x2 nodes}
We start with a small experiment within the Bordeaux site with 2 nodes on the Bordeplage compute cluster and 2 nodes on the Borderline compute cluster. We ran 30 iterations and aggregated the measured data. The measurements provide very similar metrics for all links. For such a small setting, the link connecting Bordeplage and Borderline is not a bottleneck. In agreement with this observation, the used method identified a single logical cluster containing all four nodes.

\subsubsection{32x32 nodes}
In another experiment we use 64 nodes - 32 nodes on Bordeplage, 5 nodes on Borderline and 27 nodes on Bordereau. We performed 36 BitTorrent iterations. Fig. \ref{neato:b} shows the results. It produced a perfect match to the real topology as displayed in Fig. \ref{fig:bordeaux-topology}. The two clusters Bordereau and Borderline (in circles) are merged together since they do not have a bottleneck link between them. However, the Bordeplage cluster (in diamonds) forms a different logical cluster, since it communicates to Borderline and 
Bordereau on a bottleneck 1 Gigabit link.

We also present the NMI between the specified ground truth clustering and the clustering produced by our tomography technique. Fig. \ref{fig:NMIresults} shows that after only 2 BitTorrent measurement iterations, the clustering is completely in accordance with the ground truth, and remains so during all additional iterations.
%

%

\subsection{Two site experiments}
In the next step we extend the experiments to include nodes from two sites -- Bordeaux and Toulouse. We still use 64 nodes in total -- 32 nodes per site. We described the available bandwidth within Bordeaux in the previous section. For inter-site connections between sites on Grid'5000, the optic fiber Renater network is used. This connection provides very good bandwidth (10 Gbps) for inter-site communication, but overall the inter-site bandwidth is lower than the intra-site bandwidth as described in Section \ref{ExperimentalResultsIntro}.  %
With the aggregated metric data, the clustering algorithm identifies two logical clusters, one corresponding to each of the two different sites.

Figure \ref{fig:NMIresults} shows that after 4 iterations, the clustering converges to a steady state.
However, we note that the NMI with the ground truth, while high, is imperfect -- approximately 0.7.
On investigation, we observed that this is because we have provided a ground truth within which there are 3 different partitions; for the ground truth, the network was partitioned into the Bordeaux and Toulouse sites, and then the Bordeaux site was partitioned into two separate logical clusters (as discussed in the previous section), giving a total of three separate clusters.
The best way to represent this physical setup is probably with a hierarchical representation of the clustering; however, in this work, to allow simple use of our results, we have chosen to focus on finding clusters which partition the network into a single level of clustering.

\begin{figure}
\includegraphics[width=3.50in]{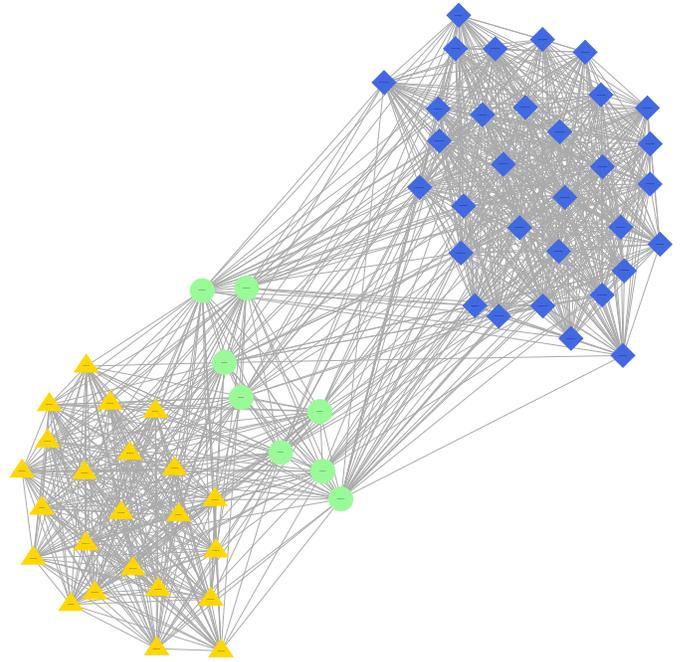}
\caption{Applying Kamada-Kawai layout to dataset `BT', a set of nodes in Bordeaux and Toulouse, using the same rendering options as for Figure \ref{neato:b}. Toulouse is represented here by diamonds; the ground truth clusters represented by circles and triangles both belong to Bordeaux. Our non-hierarchical clustering method does not recover this ground truth; it finds only two clusters, one for Toulouse and one for Bordeaux. The third ground truth cluster is distinct in the visualization, however, showing that the BitTorrent measurements do reflect it.}
\label{neato:bt}
\end{figure}

Fig. \ref{neato:bt} shows that the Kamada-Kawai layout correctly clusters Bordeaux and Toulouse, but also clusters the nodes within the Bordeaux cluster in agreement with the previous section. 
That the visualization makes visible the two separate sites within the Bordeaux cluster suggests that a future hierarchical version of our clustering step should be able to identify individual clusters within sites, at many levels, and makes clear the reason for the lower NMI in this case.

\begin{figure}
\includegraphics[width=3.50in]{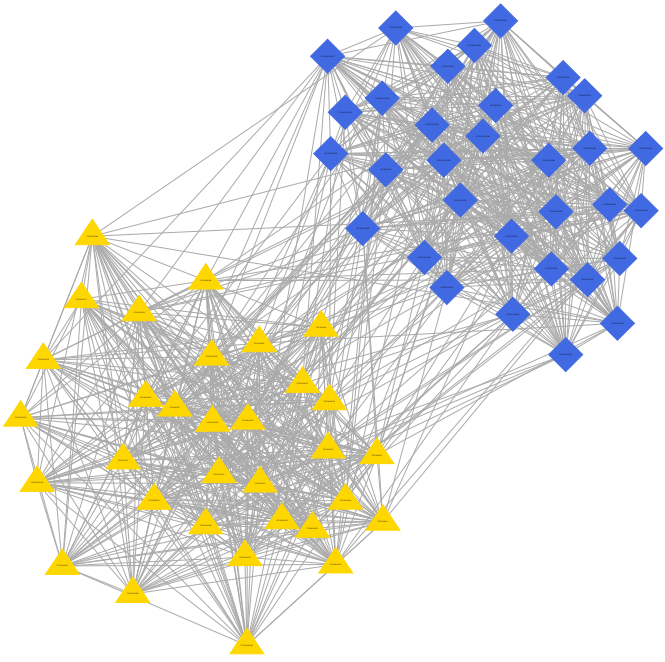}
\caption{Applying Kamada-Kawai layout to dataset `GT', a set of nodes in Grenoble and Toulouse, using the same rendering options as for Figure \ref{neato:b}.}
\label{neato:gt}
\end{figure}

In another two site experiment with two sites, we used the sites Grenoble and Toulouse, again using 64 nodes and 30 runs. Unlike Bordeaux, Grenoble and Toulouse both have a very flat Ethernet network hierarchy within them.%
As such, neither Grenoble nor Toulouse are subdivided in our ground truth.
The aggregated measurement data of our tomography method on Grenoble and Toulouse was sufficient for the clustering algorithm to identify two clusters with 100\% accuracy within the first 2 iterations (Figure \ref{fig:NMIresults}) as is clearly shown in the visualization (Figure \ref{neato:gt}).

\subsection{Three- and four-site experiments}
In the following experiments, we also use intra-site nodes which are not separated by bottlenecks within their site (e.g. in the case of Bordeaux, all nodes used are in the well connected Borderline and Bordereau physical clusters).

\begin{figure}
\includegraphics[width=3.50in]{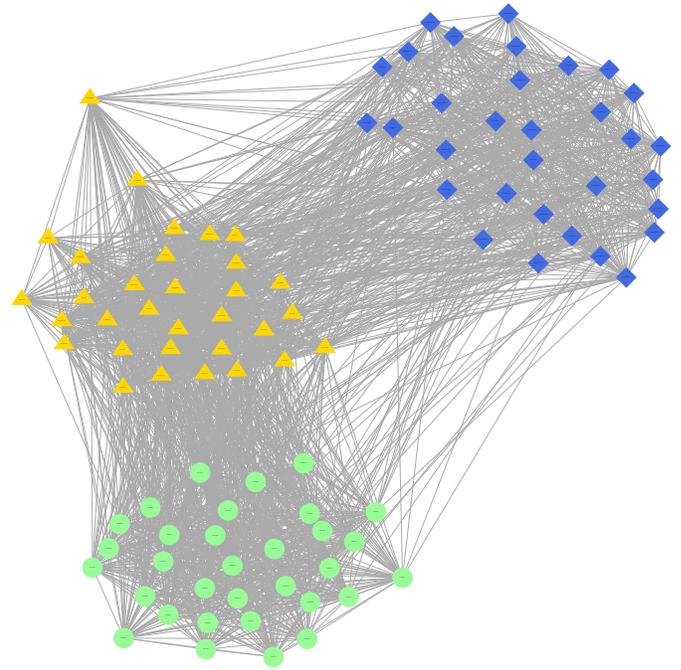}
\caption{Applying Kamada-Kawai layout to dataset `BGT', with nodes in Bordeaux, Grenoble and Toulouse, using the same rendering options as for Figure \ref{neato:b}.}
\label{neato:bgt}
\end{figure}

\begin{centering}
\begin{figure}
\includegraphics[width=3.7in]{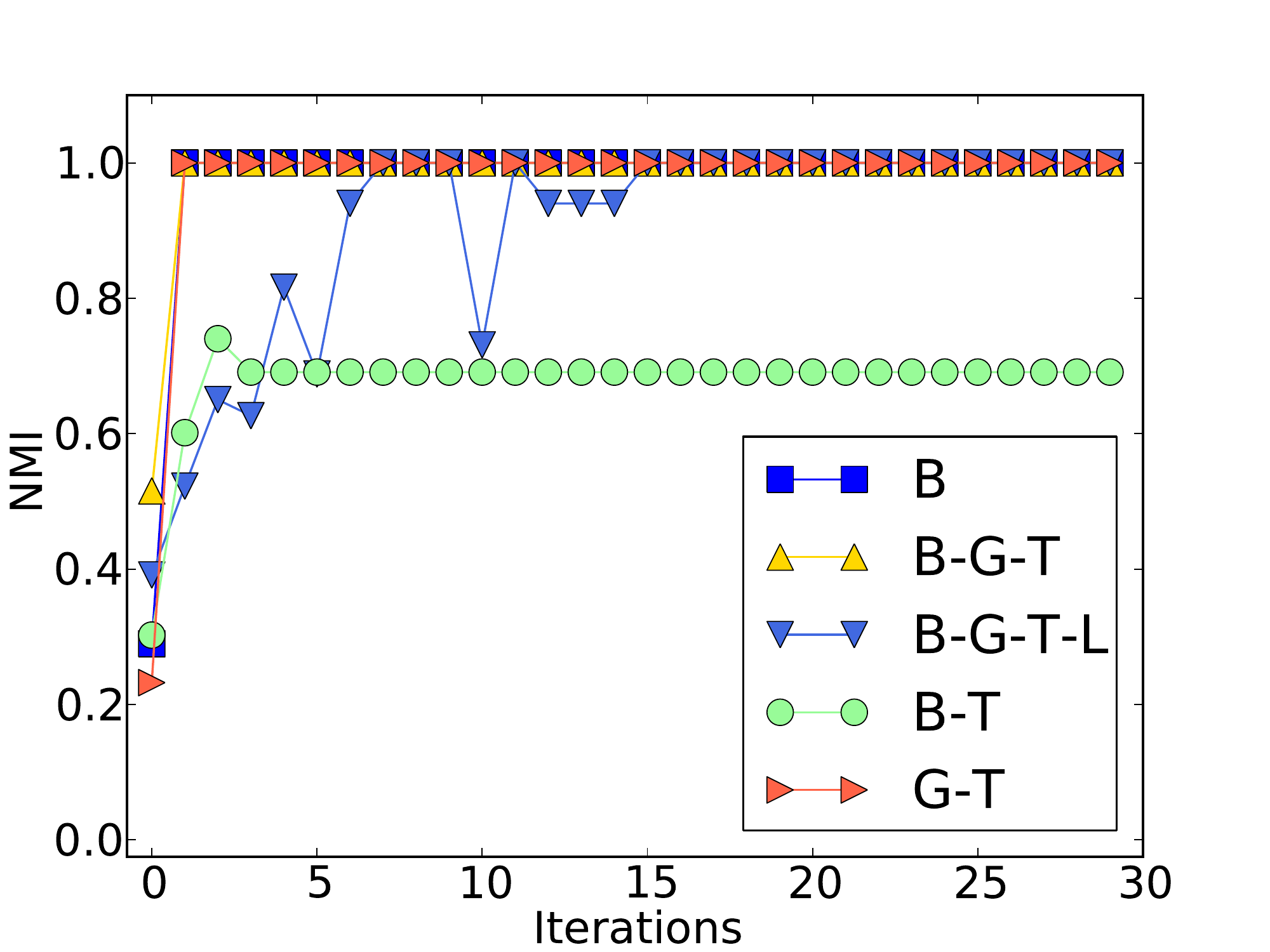}
\caption{Comparison of the clustering found using our tomography method, against the ground truth clustering provided. The results are shown in terms of Normalized Mutual Information \cite{lancichinetti2009detecting}. We observe that, in general, the NMI improves as the number of iterations performed increases, converging on a stable value. The convergence occurs quickly on the simpler topologies. The NMI frequently converges to 1 -- perfect agreement with the ground truth. In the case where NMI does not converge to 1, visualized in Figure \ref{neato:bt}, we can see that a hierarchical ground truth, and clustering approach, may improve this. Details of the topologies between the four sites used -- \textbf{B}ordeaux, \textbf{G}renoble, \textbf{T}oulouse and \textbf{L}yon -- are provided in Section \ref{ExperimentalResults}. }
\label{fig:NMIresults}
\end{figure}
\end{centering}

First, we perform a three-site experiment, using the sites Grenoble, Bordeaux and Toulouse (32 nodes per site).  Again, we perform 30 iterations, but only 2 iterations are sufficient for perfect accuracy (Fig. \ref{fig:NMIresults}) of the modularity clustering. Three clusters are identified, which are also apparent in the visualization (Fig. \ref{neato:bgt}).

\begin{figure}
\includegraphics[width=3.50in]{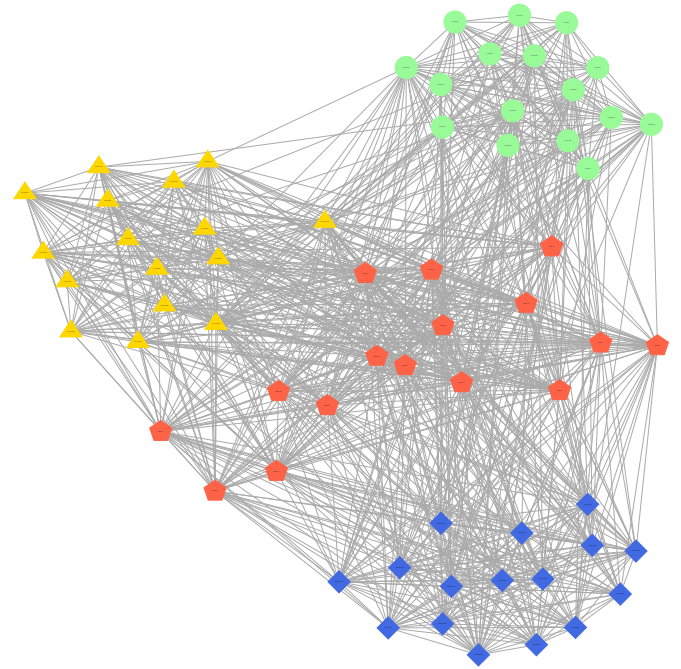}
\caption{Applying Kamada-Kawai layout to dataset `BGTL', Bordeaux, Grenoble, Toulouse and Lyon, using the same rendering options as for Figure \ref{neato:b}. The ground truth clusters in this rendering appear visually less distinct than in the other examples; however, we note that the algorithmic clustering method still manages to achieve perfect accuracy -- see Figure \ref{fig:NMIresults} below.}
\label{neato:bgtl}
\end{figure}

In the experiment which spans most sites, we use 16 nodes for each of the sites Grenoble, Bordeaux, Toulouse and Lyon. Again, we perform 30 iterations.  Modularity clustering of our BitTorrent tomography measurements correctly identifies the 4 logical clusters, which are also apparent in the visualization (Fig. \ref{neato:bgtl}). One interesting observation is that in this visualization, the central cluster of nodes represents the Lyon site, which is also positioned centrally in the star-like topology of Figure \ref{fig:renater}. 
Also interesting is that in this four-site experiment we need around 15 iterations (See Fig. \ref{fig:NMIresults}) to achieve perfect accuracy. While this is still very few, it is the largest number of iterations needed of all tested settings. This is not surprising as this is the setting with the largest number of logical clusters.

\section{Conclusion}
\label{Conclusion}

In this work, we presented a novel approach to multiple source / multiple destination network tomography. 
Instead of using traditional bandwidth measurement techniques, we counted the number of fragments exchanged in BitTorrent broadcasts. 
Even a few iterations of this approach were sufficient to allow accurate reconstruction of the logical network clusters. 

The reconstruction was done using a modern network clustering algorithm -- modularity based clustering. 
Our experimental results show that we can reliably find clusters with bandwidth tomography. 
Our approach is much more efficient at revealing network properties which appear under intense collective communication than existing methods. 
Existing methods would take hours or days to uncover these details; our approach requires only a few minutes, and achieves high accuracy.

We also evaluated the number of BitTorrent broadcasts needed for various settings to achieve this accuracy. 
Correct clustering within a single site needed only a small number of iterations; whereas around 15 iterations are needed for our most complex experiment, taking only a few minutes, and running on a larger empirical network than studied in related bandwidth tomography literature.

Our approach correctly identified communication bottleneck links in physical clusters by placing the nodes communicating across the bottleneck link in different logical clusters - a significant result. 
It also separated nodes in different sites into different logical clusters due to the lower inter-site bandwidth.
 
The efficiency of our approach makes it useful for applications relying on bulk data transfer -- e.g. applications performing all-to-all operations -- across complex and heterogeneous networks of computers.

This automatic efficient measurement is also particularly suitable for overlay networks, or networks of virtual machines, which may have a dynamically altering underlying topology.

Our results were robust for all settings tested.

\subsection*{Future Work}
We have seen promising results with this technique.
A major advantage of the technique is that all parts of it -- the BitTorrent based measurement technique, and the clustering algorithm -- are designed to scale to large networks.
As we have shown that the clusterings found accurately uncover network structure in empirical networks for which ground truth is available, future work should integrate this tomography method into existing parallel computation libraries, and measure the performance increase gained on large networks.
This would allow us further evaluate the effectiveness of this method in an application setting.

In this work, to keep our method compatible with existing software libraries, we have worked only with ground truths that are a single partitioning of the network into disjoint non-overlapping clusters.
However, both the network clustering algorithm used, and the NMI evaluation method, extend to overlapping multi-level hierarchical clusterings.
Extending our measurement technique and investigating the performance of the tomography approach on such hierarchical datasets would be valuable.

\section*{Acknowledgment}
This publication has emanated from research conducted with the financial support of Science Foundation Ireland under Grant Number 08/IN.1/I2054 and 08/SRC/I1407.

Experiments presented in this paper were carried out using the Grid'5000 experimental testbed, being developed under the INRIA ALADDIN development action with support from CNRS, RENATER and several Universities as well as other funding bodies (see https://www.grid5000.fr).

Special thanks to Sebastien Badia for clarifying the network topology within the Bordeaux site of Grid'5000.

\bibliographystyle{IEEEtran}
\bibliography{reference-list-jabref}
\end{document}